\documentclass[journal, 10pt, twoside]{IEEEtran}

\usepackage[american]{babel}
\usepackage{tabularx}
\usepackage{multirow}
\usepackage{booktabs}
\usepackage{amsmath}
\usepackage{amssymb}
\usepackage{cite}
\usepackage{algorithm}
\usepackage{algorithmicx}
\usepackage{algpseudocode}
\usepackage{graphicx}

\usepackage{listings}
\usepackage{placeins}
\usepackage{xspace}
\usepackage{color}
\usepackage{colortbl}

\usepackage{subcaption}

\usepackage[figuresright]{rotating}

\newcommand{\ourprop}{DADO\xspace}
\newcommand{\eourprop}{Distributed Application Deployment Optimization\xspace}

\definecolor{Linen}{rgb}{0.9803,0.9411,0.9019} 
\definecolor{White}{rgb}{1,1,1}
\definecolor{Coral}{rgb}{1,0.4980,0.3137}
\definecolor{Grayblue}{rgb}{0.9411,0.9411,0.9803}
\definecolor{DarkLinen}{rgb}{0.729,0.7176,0.635}
\definecolor{Green}{rgb}{0,0.5,0}
\definecolor{Red}{rgb}{0.5,0,0}

\algdef{SE}[DOWHILE]{Do}{doWhile}{\algorithmicdo}[1]{\algorithmicwhile\ #1}%
\graphicspath{ {figures/} }

\usepackage{scalerel}
\usepackage{tikz}
\usetikzlibrary{svg.path}

\definecolor{orcidlogocol}{HTML}{A6CE39}
\tikzset{
    orcidlogo/.pic={
        \fill[orcidlogocol] svg{M256,128c0,70.7-57.3,128-128,128C57.3,256,0,198.7,0,128C0,57.3,57.3,0,128,0C198.7,0,256,57.3,256,128z};
        \fill[white] svg{M86.3,186.2H70.9V79.1h15.4v48.4V186.2z}
        svg{M108.9,79.1h41.6c39.6,0,57,28.3,57,53.6c0,27.5-21.5,53.6-56.8,53.6h-41.8V79.1z M124.3,172.4h24.5c34.9,0,42.9-26.5,42.9-39.7c0-21.5-13.7-39.7-43.7-39.7h-23.7V172.4z}
        svg{M88.7,56.8c0,5.5-4.5,10.1-10.1,10.1c-5.6,0-10.1-4.6-10.1-10.1c0-5.6,4.5-10.1,10.1-10.1C84.2,46.7,88.7,51.3,88.7,56.8z};
    }
}

\newcommand\orcidicon[1]{\href{https://orcid.org/#1}{\mbox{\scalerel*{
                \begin{tikzpicture}[yscale=-1,transform shape]
                \pic{orcidlogo};
                \end{tikzpicture}
            }{|}}}}

\usepackage{hyperref}

\begin{document}

\title{Optimizing Response Time in SDN-Edge Environments for Time-Strict IoT Applications}

\author{Juan~Luis~Herrera$^{\textsuperscript{\orcidicon{0000-0002-2280-2878}}}$\,, Jaime~Gal{\'a}n-Jim{\'e}nez$^{\textsuperscript{\orcidicon{0000-0002-5476-7130}}}$\,, Javier~Berrocal$^{\textsuperscript{\orcidicon{0000-0002-1007-2134}}}$,~\IEEEmembership{Member,~IEEE}, and\\ Juan~Manuel~Murillo$^{\textsuperscript{\orcidicon{0000-0003-4961-4030}}}$,~\IEEEmembership{Member,~IEEE}%
\thanks{Manuscript received January 00, 0000; revised January 00, 0000; accepted January 00, 0000. Date of publication January 00, 0000; date of current version January 00, 0000. This work has been partially funded by the project RTI2018-094591-B-I00 (MCI/AEI/FEDER,UE), the 4IE+ Project (0499-4IE-PLUS-4-E) funded by the Interreg V-A Espa\~na-Portugal (POCTEP) 2014-2020 program, by the Department of Economy and Infrastructure of the Government of Extremadura (GR18112, IB18030), and by the European Regional Development Fund. \emph{(Corresponding author: Juan Luis Herrera.)}}%
\thanks{The authors are with the Department of Computer Science and Communications Engineering, University of Extremadura, Spain (e-mail: jlherrerag@unex.es).}%
\thanks{Digital Object Identifier 00.000/JIOT.0000.0000000}
}

\markboth{IEEE Internet of Things Journal, Vol. 0, No. 0, January 0000}{Herrera \MakeLowercase{\textit{et al.}}: \ourprop: Optimizing the Quality of Service of IoT Applications in SDN-Edge Environments}

\IEEEpubid{0000--0000/00\$00.00 ~\copyright~2020 IEEE}

\maketitle

\begin{abstract}
WARNING - This version of the paper is incomplete and outdated. You can find the newer version at \url{https://doi.org/10.1109/JIOT.2021.3077992}

The rise of the Internet of Things (IoT) has opened new research lines that focus on applying IoT applications to domains further beyond basic user-grade applications, such as Industry or Healthcare. These domains demand a very high Quality of Service (QoS), mainly a very short response time. In order to meet these demands, some works are evaluating how to modularize and deploy IoT applications in different nodes of the infrastructure (edge, fog, cloud), as well as how to place the network controllers, since these decisions affect the response time of the application. Some works in the literature have approached this problem by providing separate plans for deployment and placing of controllers. However, this approach makes sub-optimal decisions, that complicate guaranteeing the demanded response time. To guarantee an optimal response time, it is crucial to solve the problem in a single effort that considers both, the networking and computing dimensions. In this work, we analyze the influences between the response time of computing and networking in edge computing environments with SDN networks, merging both optimization efforts into a single one and proposing a solution to the joint problem. Our evaluation shows that our proposal can shorten response time by up to 28.97\%.
\end{abstract}


\section{Introduction} \label{sec:introduction}

\IEEEpubidadjcol

\IEEEPARstart{T}{he popularity} of IoT devices for the general public have made them ubiquitous. We live surrounded by everyday objects (\emph{things}) that are connected to the Internet and run IoT applications: programs that are able to interact with the real world through IoT devices, getting inputs from it through their sensors and changing it through their actuators. This makes IoT applications interesting for different domains, both general purpose domains such as domotics and intensive domains such as Industry or Healthcare.

To run these IoT applications, the cloud computing paradigm is the most common option\cite{Singh2016}. In cloud computing, a set of powerful servers, generally far away from the IoT devices, do the processing, while IoT devices only have to send their requests to these servers. While a cloud-centric architecture is enough for user-grade, general purpose applications\cite{Yousefpour2019}, a purely cloud-centric architecture may not be enough to meet the QoS requirements of complex and intensive applications. Industrial Internet of Things (IIoT) applications such as Factory Automation need very low response times\cite{Xu2018}, while Internet of Medical Things (IoMT) applications can be very time-sensitive as well on the execution of artificial intelligence models\cite{Greco2020}. Therefore, the objective of this paper is to minimize the response time of intensive IoT applications such as IIoT or IoMT applications. Cloud computing servers are often far away from IoT devices, and may thus have latencies that may complicate meeting a short enough response time.

\IEEEpubidadjcol

It is for this reason that other paradigms, such as fog computing or mist computing, which can be referred to under the umbrella term \emph{edge computing}\cite{Baktir2017}, are emerging to support applications with strict response time requirements \cite{Yousefpour2019}. Edge computing takes advantage of the computation capabilities of different nodes closer to end devices, as well as the devices themselves. To do so, it executes different parts of IoT applications in them. With this approach, it is simpler to get shorter response times than by using pure cloud computing infrastructures. Therefore, in edge computing environments, which nodes to use and which node should host which parts of the application are decisions that have to be taken, and have an impact on the provided response time\cite{Bellavista2019}. The problem of taking optimal decisions, and thus distributing computation optimally, is known as the Decentralized Computation Distribution Problem (DCDP) \cite{Choudhury2019}.

A key element of the DCDP, as well as one of the main motivations behind edge computing paradigms, is network latency\cite{Bellavista2019}: the latency from IoT devices to nearby edge devices is smaller than to the cloud, hence execution time can be shortened. This means that minimizing network latency, through techniques such as routing optimization\cite{Wang2008}, is key for correctly solving the DCDP. Software-Defined Networking (SDN), a network paradigm that allows networks to be programmed by means of SDN controllers, allows routing optimization to be performed in a programmable way\cite{Baktir2017}, making this latency optimization scalable and flexible as new devices are added to the infrastructure. SDN lets network controllers be notified when new devices are added through discovery protocols\cite{Baktir2017}, as well as it allows network controllers to monitor networking and computing devices\cite{Baktir2017}, gathering performance information from the infrastructure that can be used to solve the DCDP.

While the control that SDN provides allows for latency between computing devices to be optimized to a some extent, the control latency of the SDN network can also be optimized. SDN networks rely on controllers to work, and therefore, the latency between SDN switches and controllers affects the latency between any two devices in the SDN network\cite{Das2019}. This implies that, if controllers are placed in such a way that this latency is minimized, the latency between devices in the network is minimized as well. The problem of placing the controller optimally is known on research works as the SDN Controller Placement Problem (CPP)\cite{Das2019}.

Several works in literature have studied both the DCDP and the CPP as separate problems~\cite{Choudhury2019,Carrega2017,Sun2018,Brogi2020,Lv2019,Zhang2018,ulHuque2015,Bari2013,Singh2020,Das2019}, providing partial solutions that only consider one dimension. The DCDP solutions assume that the network is an static entity that provides a certain latency, while the CPP solutions assume that the traffic flows do not change depending on the latency achieved by the network. However, solving each of these problems separately may not result in a response time optimization that is enough for IoT applications with strict response time requirements. The decision of assigning a service to a node in a certain layer should take into account the network latency to be optimal, which depends on controller placement; and the decision of placing the controller should consider the steering of traffic flows throughout the network, which depends on which nodes are requesting services and which nodes are providing them. Therefore, these dimensions affect each other. To fully optimize the response time of IIoT, IoMT and other intensive IoT applications, both problems should be solved together so their mutual influences and trade-offs can be taken into account, merging into a single new problem. We define the resulting combined problem as the Edge-SDN Deployment Problem. To the best of our knowledge, no prior work has addressed the Edge-SDN Deployment Problem. The main contributions of this work are:

\begin{itemize}
    \item A study of the relationship between the DCDP and the CPP in intensive IoT environments.
    \item The definition of the Edge-SDN Deployment Problem.
    \item The formalization of the Edge-SDN Deployment Problem.
    \item The proposal of a framework named \eourprop (\ourprop) to solve the Edge-SDN Deployment Problem.
    \item An evaluation of \ourprop in an IIoT scenario.
\end{itemize}

This paper is structured as follows: Sec. \ref{sec:motivation} motivates the combination of the CPP and the DCDP into the Edge-SDN Deployment Problem through an illustrative IIoT example. Sec. \ref{sec:proposal} presents the proposal for \ourprop, a solution to the Edge-SDN Deployment Problem. Sec. \ref{sec:evaluation} evaluates \ourprop on a setup based on the previously presented example. Sec. \ref{sec:related} presents related work. Finally, Sec. \ref{sec:conclusions} concludes the paper and highlights future challenges.
\section{Motivation} \label{sec:motivation}

To illustrate the importance of combining the CPP and the DCDP, a particular case study scenario is presented in this section.

\subsection{Scenario: edge IIoT factory} \label{subsec:scenario}

The scenario presented in this section is based on the environment proposed in \cite{Sun2018}, since this work provides not only an IIoT-based edge computing scenario, but also enough details about the infrastructure so it is possible to apply and evaluate our solution, DADO. In this scenario, an edge infrastructure based on an SDN network is deployed in a factory to transform it into a Cyber-Physical smart factory by leveraging IIoT\cite{Xu2018}. An IIoT device is installed in each robot: initially, only 10 robots are part of the smart factory, but this system is expected to grow over time if the company decides to further invest into the transformation. At the same time, 10 edge servers are placed in the same factory to provide the IIoT devices with services. Each edge server has a 800 MHz CPU and 1 GB of RAM memory\cite{Sun2018}, and is directly connected to an SDN switch.

\begin{figure}[tb]
	\centering
	\includegraphics[width=0.9\columnwidth]{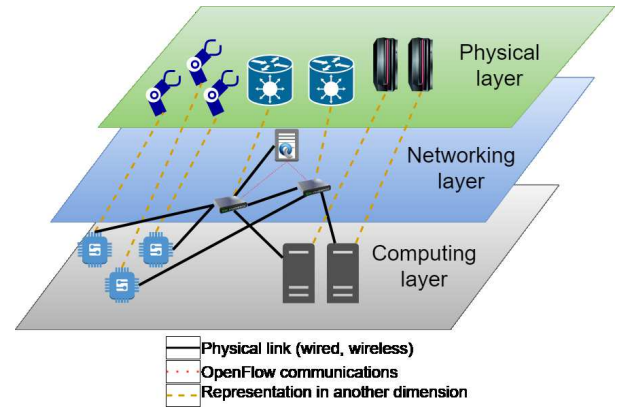}
	\caption{Presented scenario in the three involved layers.}
	\label{fig:completecs}
\end{figure}

In this factory, SDN is leveraged to provide service discovery due to its properties as a modular, independent and transparent solution\cite{Baktir2017}. Since SDN networks need at least one controller, the factory uses the classic SDN control model and co-locates them with SDN switches \cite{Das2019}. Fig. \ref{fig:completecs} shows this Cyber-Physical IIoT System\cite{Xu2018}, divided into three layers: the physical layer embodies the \emph{physical} part of the system, while the \emph{cyber} part is divided into two: the networking layer, which contains the SDN switches and controllers; and the computing layer, which contains the IIoT devices and edge servers.

\begin{figure*}[h!]
	\begin{subfigure}[b]{0.5\textwidth}
	    \includegraphics*[width=\textwidth]{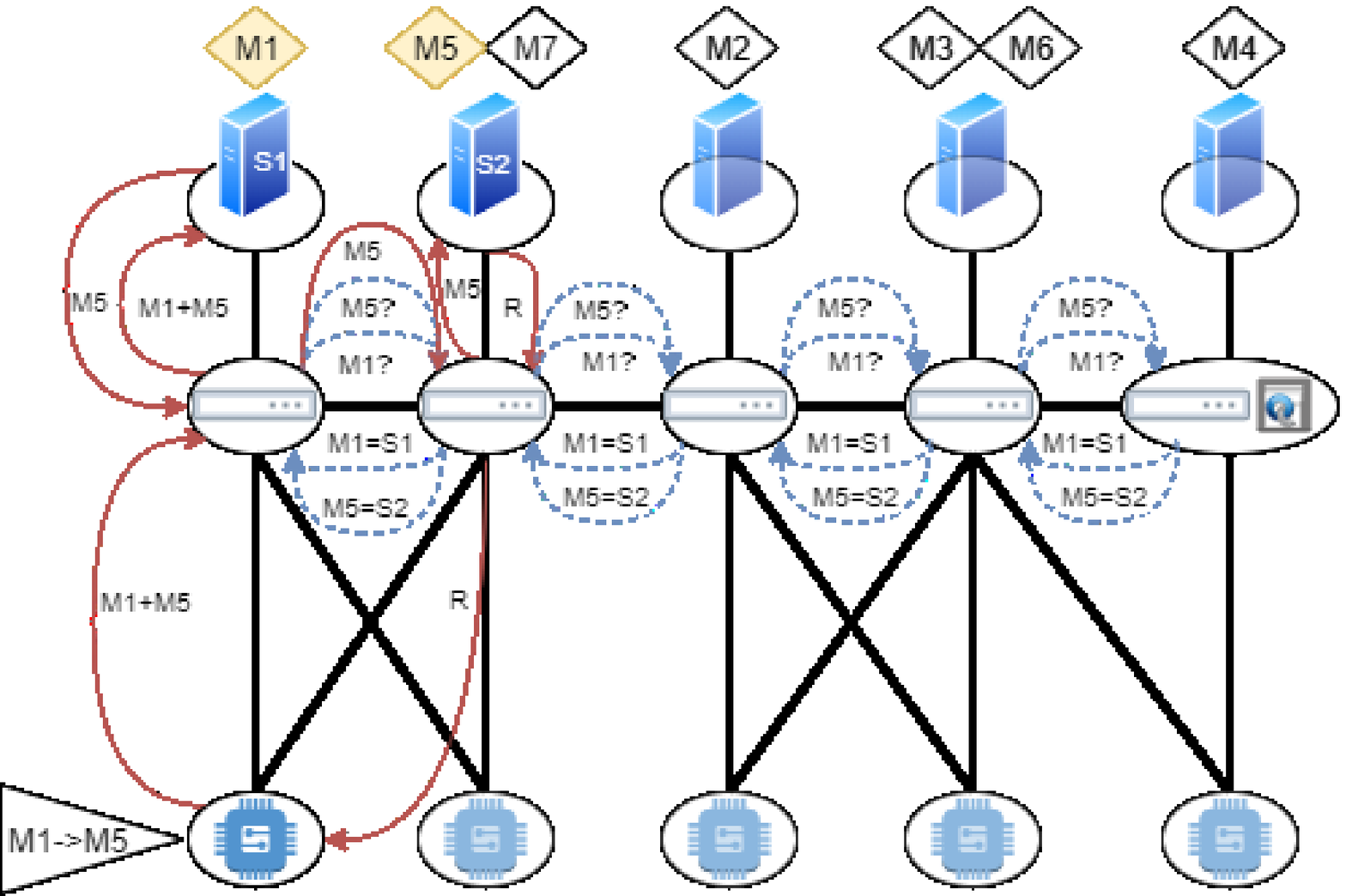}
	    \caption{Edge deployment, sub-optimal controller placement}
	    \label{subfig:suboptimal}
	\end{subfigure}
	\hfill
	\begin{subfigure}[b]{0.5\textwidth}
	    \includegraphics*[width=0.9\textwidth]{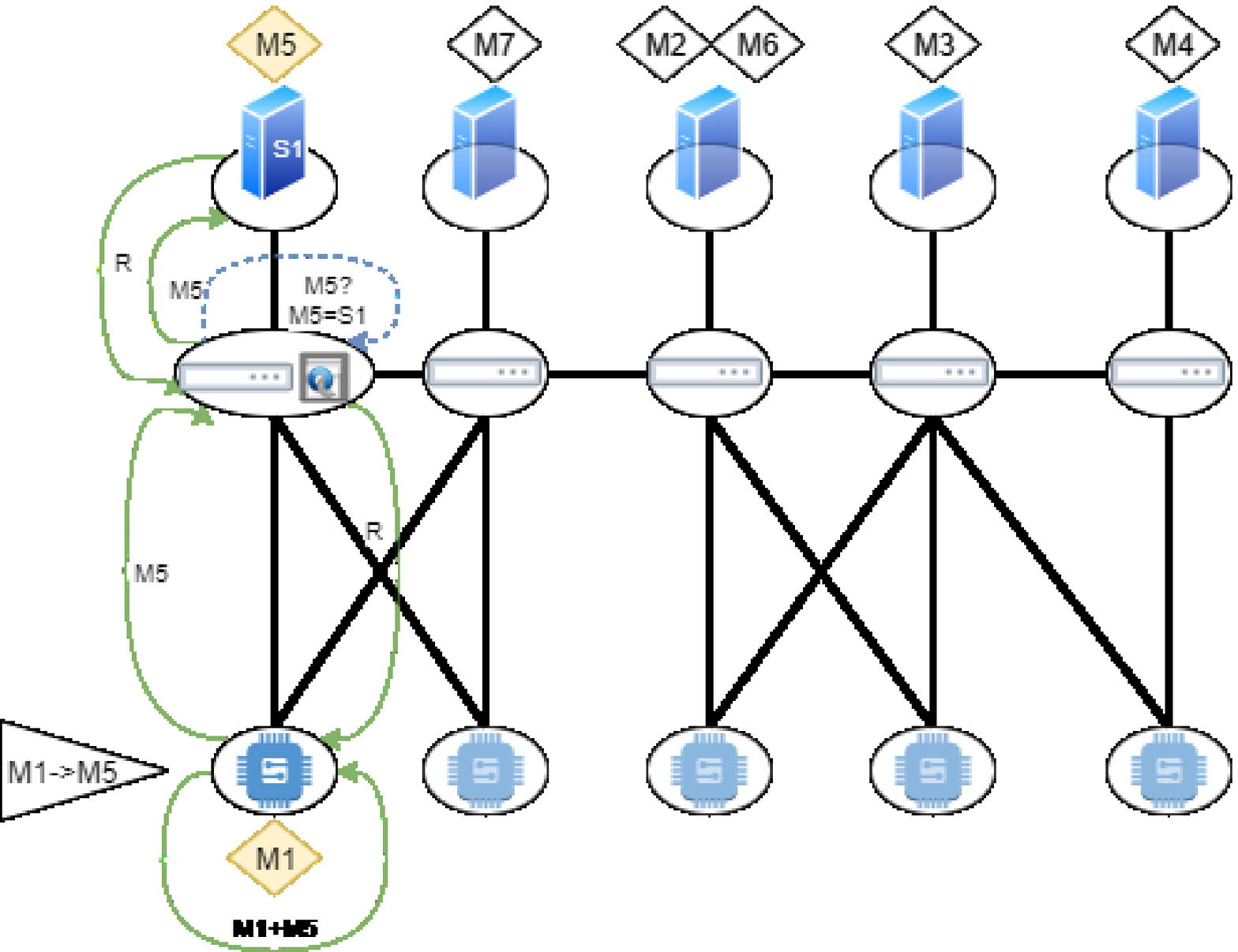}
	    \caption{Mist and edge deployment, optimal controller placement}
	    \label{subfig:optimal}
	\end{subfigure}
	\caption{Workflow execution with two different deployment strategies.}
	\label{fig:strategies}
\end{figure*}

In this infrastructure, an IIoT application is to be deployed to continuously monitor and manage the smart factory, gathering the status of robots through the sensors connected to IIoT devices and processing them in edge servers to provide according commands. This application is designed using a Micro Services Architecture (MSA), and is therefore composed of different independent services that perform a certain type of processing, whose functions can be requested separately or can be combined through workflows. Each microservice takes 100, 500 or 1000 MCycles to run\cite{Sun2018}, according to its computational complexity. Each workflow comprises the execution of a certain functionality, chaining between 1 and 6 microservices, depending on the amount of them required to perform the functionality. Response time is the combination of execution time (i.e. the time it takes to execute the microservices of a functionality) and latency (i.e. the time it takes to communicate from a computer to another computer in a network).

Fig. \ref{fig:strategies} shows the influence of the location of microservices and SDN controllers in response time: a sample workflow is shown, which performs a command request functionality by chaining a microservice that aggregates information from multiple sensors (M1) with another microservice that analyzes the aggregated information to issue a corresponding command (M5). In this figure, the box with the OpenFlow logo represents the SDN controller, dashed lines represent SDN control messages and solid lines represent application messages. In Fig. \ref{subfig:suboptimal}, M1 is located in edge server S1 and M5 is located in edge server S2. As the leftmost IIoT requests a command, its message is sent to the SDN network, with M1 as the destination address. The SDN switch is not aware of the location of M1, so it asks the SDN controller where it can be found through an OpenFlow packet-in message to perform service discovery. The SDN controller answers with a packet-out message, and the switch routes the message to S1. Once the data is aggregated on S1, the message requesting for M5 is sent to the network to analyze it. Again, the SDN switch must request service discovery to find M5. As the packet-out message arrives, the message is routed to S2. Finally, S2 issues the command, sending it back to the IIoT device through the network. In this exchange, messages had to be sent 23 times through the network, each new sending increasing the delay due to latency. Fig. \ref{subfig:optimal} shows a different strategy, in which the IIoT device itself aggregates the information and the SDN controller is placed closer. Because of these changes, the IIoT device only needs to request M5. Since the SDN controller is placed on the same SDN switch the message is sent to, service discovery is performed locally with minimal latency. S1 analyzes the aggregated information and sends back the command as previously. This new strategy changes the number of message sendings to 4, reducing the overall execution time. Therefore, to optimally deploy this application, two decisions must be made: i) In which host to execute each microservice in the architecture, and ii) In which switch or switches to place SDN controllers.

\subsection{DCDP: Placing microservices} \label{subsec:cdp}

The solution to the DCDP aims to make the first decision: selecting which microservices should run in the IIoT devices themselves and which microservices should run in edge servers. As seen in Fig. \ref{fig:strategies}, while IIoT devices are not as powerful as edge servers, executing some microservices in them, especially those that are lightweight and requested very often, allows parts of the workflows to be executed locally, completely ignoring network latency. The difference is on how much faster can these services be executed when they are placed in the edge, and how large the latency between IIoT devices and edge servers is. If the latency to the edge servers is larger than the difference in execution time, it is worth it to execute microservices locally; however, if latency is smaller than this difference, response time is shorter if they run on edge servers. Therefore, the choices of the DCDP are inherently related to network latency. Thus, to optimize response time through the DCDP, network latency should be optimized first.

\subsection{CPP: Placing SDN controllers} \label{subsec:cpp}

The problem the CPP tries to solve is related to the other decision in the scenario: where to place SDN controllers. The placement of the SDN controller plays a key role on control latency and, by extension, on the overall network latency\cite{Das2019}. Controller placement is very related to the network topology, but also to how traffic flows are steered through the network\cite{Singh2020}: while topology-wise a node may seem optimal for controller placement, it may not be optimal if that zone of the network is not very used. This case was shown in Fig. \ref{subfig:optimal}, where the controller is placed on the leftmost part of the network because traffic flows are steered through said zone.

In this scenario, traffic is generated by the IIoT application when a workflow cannot be fully executed locally: in these cases, one host sends a message with the input data of the microservice to the host that executes it. Once executed, the output data of the microservice is sent back. The execution of M5 in Fig. \ref{subfig:optimal} shows a graphical example of this. These messages generate two traffic flows: one to sends the input data, and another one to send the response. Thus, the decisions taken during the DCDP may affect the CPP: the choice to execute a microservice locally removes traffic flows, which may make its area less suitable for controller placement; while the choice to execute a microservice remotely adds traffic flows, which may make that area more suitable for controller placement. Sub-optimal decisions in the DCDP may generate sub-optimal decisions in the CPP as well. Thus, to optimize the response time through the CPP, microservices should be placed in a way that optimizes response time.

This generates a bootstrapping problem: in order to optimize the network layer through the CPP, the computing layer should be optimized first through the DCDP; but to optimize the computing layer through the DCDP, the network should be optimized first through the CPP. Considering them separately implies possibly making sub-optimal decisions because of these behavior differences, and thus arriving at a sub-optimal solution, such as the one shown in Fig. \ref{subfig:suboptimal}. To avoid the possibility of getting a sub-optimal response time for the IIoT application, both computing and networking should be optimized at the same time to avoid this bootstrapping problem.
\section{\ourprop framework} \label{sec:proposal}

In this section, the \ourprop framework is presented. \ourprop is a framework that solves the Edge-SDN Deployment Problem. To do so, \ourprop takes as an input statistics such as the computational load of microservices, the computational power of any edge, fog, cloud or mist servers; the latency and capacity of links, the workflows used by the requested functionalities or the network topology. This infrastructure description, that can be obtained by leveraging SDN\cite{Baktir2017}, allows \ourprop to solve the Edge-SDN Deployment Problem and to provide useful information. Concretely, \ourprop provides as an output information about the placement of network controllers, the placement of microservices in hosts and the routes taken by traffic in the network, optimizing all of them to provide the optimal response time. In order to solve the Edge-SDN Deployment Problem, \ourprop uses at its core a Mixed Integer Linear Programming (MILP) formulation, as presented in the following.

\ourprop aims at optimizing response time in mist, edge and fog architectures, as well as in hybrid ones, for IoT applications based on an MSA. Thus, the infrastructure will contain SDN switches, IoT devices, edge servers, fog nodes and cloud nodes. IoT devices, as well as edge servers and fog and cloud nodes, are able to execute parts of the logic of the application, generate traffic and consume this traffic; they are \emph{hosts}. SDN switches have a completely different behaviour, forwarding traffic by applying routes calculated by their assigned SDN controller. Therefore, let the infrastructure be represented as a graph $G=\{V, L\}$. Let $H$ be a set of hosts and $S$ be a set of SDN switches so that the set of vertices $V=H\cup S$. Let $L$ be the links that connect the different elements of the infrastructure.

Not all of these hosts have the same capabilities. Generally, edge severs and fog nodes are more powerful than IoT devices, with cloud servers being the most powerful. This power can be represented by its speed executing microservices, as well as by the maximum amount of services it can execute. Thus, let a host $h\in H$ be a tuple $h=<P_h, r_h>$, with $P_h$ being the computational power of the host (measured as its clock speed in Hz) and $r_h$ being the amount of RAM memory of the host, measured in bytes. If other applications or services are running at such host, then $P_h$ is the computational power of the host that is not being used for other application or services (i.e. they can be used by the IoT application) and $r_h$ is the remaining free RAM memory of the host.

In this infrastructure, links have two essential limitations. Firstly, links have an impact on latency, since sending data over them is not instantaneous. Secondly, links do not have infinite capacity, and therefore cannot be used to transfer an unlimited amount of data. Thus, let a link $l_{ij}\in L$, with $i$ being the source of the link and $j$ being its destination, be a tuple $l_{ij}=<\delta_{ij}, \theta_{ij}>$, with $\delta_{ij}$ being the latency of the link in seconds, and $\theta_{ij}$ being the maximum capacity of the link in bytes per second. If the link is also being used to transmit data that is not related to the application that is being optimized, $\theta_{ij}$ only references the capacity that is not in use by other traffic.

The IoT applications \ourprop supports have an MSA, and can be seen effectively as a set of independent microservices. We therefore have a set of microservices $M$, with each microservice $m\in M$ being a tuple $m=<\xi_m, I_m, O_m, r_m>$, with $\xi_m$ as the workload of executing the microservice (measured as the number of CPU cycles it requires to fully execute), $I_m$ being the size of the input data for the microservice in bytes, $O_m$ being the size of the output data of the microservice in bytes and $r_m$ being the amount of RAM memory the microservice requires in bytes.

The execution model of \ourprop is based on workflows. When an IoT device requests a certain functionality, this functionality is provided by a workflow of one or more microservices.

Where these microservices run depends on the solution \ourprop generates. If the first microservice, or two consecutive microservices, are ran in the same host, then the network is not used. If the microservice does not run in said device, then the SDN network is used to route the request to a host that runs the service. Let $W$ be a set of workflows, with each workflow $w\in W$ being an ordered set of tuples $w=\{c_1, c_2, ..., c_{|w|}\}$. Each tuple $c_i$ must have the exact same format and values as one of the microservices in $M$, since each of these tuples represents the microservices that are chained through the workflow to perform a functionality. Therefore, with a slight abuse of notation, we can say that $w=\{c_1, c_2, ..., c_{|w|}\}; c_i\in M\forall i \in [1, |w|]$. Let also $WS(w,h)$ be a binary function that is 1 if workflow $w$ is started by host $h$ and 0 otherwise. To execute the workflow, the data would have to flow from whatever host starts the workflow to a host that executes $c_1$, from there to the one that executes $c_2$, and so on.

We also propose the usage of the classic SDN control model\cite{Das2019}. Each controller can be co-located with an SDN switch in the network $s\in S$ (i.e. the controller and the SDN switch are in the same place). Each SDN switch is \emph{mapped} to a controller, so that it communicates with it through in-band traffic. Thus, let $\psi$ be the maximum amount of SDN controllers to be placed and let $\Omega$ be the size of the control packets that are sent from each SDN switch to the controller.

After these parameters have been defined, we have to set different different decision variables that must be changed to optimize the deployment.

In the computing plane, let $z$ be a three-dimensional binary matrix, in which $z_{hc_a}^{w}$ is $1$ if host $h$ is running the microservice indexed as $c_a$ of workflow $w$ and $0$ otherwise. This allows \ourprop to locate microservices in certain hosts. Let $f$ be a five-dimensional binary matrix, in which $f_{ij}^{hwc_a}$ is $1$ if the traffic host $h$ generates as a consequence of the microservice indexed $c_a$ of workflow $w$ is routed through the link $l_{ij}$ and $0$ otherwise. This allows \ourprop to route the input and intermediate output data of microservices through the network. Let $f'$ be a four-dimensional binary matrix, in which $f_{ij}^{\prime hw}$ is $1$ if the traffic host $h$ generates as the response for workflow $w$ is routed through the link $l_{ij}$ and $0$ otherwise. This lets \ourprop route the final output data of microservices.

In the networking plane, let $x$ be a binary vector, in which $x_s$ is $1$ if an SDN controller is set up on switch $s$ and $0$ otherwise. This lets \ourprop place SDN controllers. Let $y$ be a binary matrix, in which $y_{ss'}$ is $1$ if SDN switch $s$ is mapped to controller $s'$ and $0$ otherwise. This allows \ourprop to map SDN controllers and switches. Let $cf$ be a three-dimensional binary matrix, in which $cf_{ij}^s$ is $1$ if the control traffic switch $s$ generates is routed through the link $l_{ij}$ and $0$ otherwise. This lets \ourprop route the control data between SDN switches and their mapped controllers.

We also need to establish constraints that determine which values are allowed for each variable under different conditions and how changing these values affects the overall deployment. First, so far we assume that a microservice for a certain workflow can only be executed in a single host.

\begin{equation}
    \sum_{h\in H}z_{hc_a}^{w} = 1; \forall w\in W, a\in [1, |w|]
    \label{eq:one}
\end{equation}

Then, each host cannot run unlimited microservices, but only as many as its memory allows.

\begin{equation}
    \sum_{w\in W}\sum_{a = 1}^{|w|}z_{hc_a}^{w}r_{c_a} \le r_h; \forall h\in H
    \label{eq:two}
\end{equation}

It is not possible to have more controllers than the maximum amount.

\begin{equation}
    \sum_{s\in S}x_s \le \psi
    \label{eq:four}
\end{equation}

A switch can only be mapped to one controller at a time.

\begin{equation}
    \sum_{s'\in S}y_{ss'} = 1; \forall s\in S
    \label{eq:five}
\end{equation}

It can also only be mapped to a controller if said controller is actually placed.

\begin{equation}
    y_{ss'} \le x_{s'}; \forall s, s'\in S 
    \label{eq:six}
\end{equation}

Flow variables should be controlled on an aggregate, according to the classic flow constraints. When we account for microservice data, we need to consider three different cases: the first microservice of the workflow ($c_1$), the response of the workflow, and the general case for the rest. In the case of $c_1$, it can be stated that: i) Traffic is generated only by the host that starts the workflow, unless $c_1$ is mapped to the same host (in that case, it will be executed locally); and ii) Traffic is consumed by the host that has $c_1$ mapped, unless it is the same host that starts the workflow. Formally:

\small
\begin{equation}
    \begin{split}
    \sum_{j\in V} f_{ij}^{hwc_1}-f_{ji}^{hwc_1} & =\begin{cases}
    0 & \text{if }i\in S \\
    WS(w,h)(1-z_{ic_1}^{w}) &\text{if }i = h\\
    -WS(w,h)z_{ic_1}^{w} &\text{otherwise.} \\ 
    \end{cases} \\
    & \forall i\in V, h\in H, w\in W
    \end{split}
    \label{eq:seven}
\end{equation}
\normalsize

In the case of the response, we can state that: i) Traffic is generated only by the host that has the last microservice mapped, unless it is mapped to the host that started the workflow; and ii) Traffic is consumed by the host that started the workflow, unless it has the last microservice mapped. Formally:

\small
\begin{equation}
    \begin{split}
    \sum_{j\in V} f_{ij}^{\prime hw}-f_{ji}^{\prime hw} & =\begin{cases}
    0 & \text{if }i\in S \\
    z_{hc_{|w|}}^{w}(1-WS(w,i)) &\text{if }i = h\\
    -z_{hc_{|w|}}^{w}WS(w,i) &\text{otherwise.} \\ 
    \end{cases} \\ 
    & \forall i\in V, h\in H, w\in W
    \end{split}
    \label{eq:eight}
\end{equation}
\normalsize

We can derive a general case. Traffic is generated by the host that has the previous microservice mapped, as long as the current microservice is not mapped to it; and it is consumed by the host that has the current microservice mapped, as long as it does not have the previous microservice mapped.

\small
\begin{equation}
    \begin{split}
    \sum_{j\in V} f_{ij}^{hwc_a}-&f_{ji}^{hwc_a}  =\begin{cases}
    0 & \text{if }i\in S \\
    z_{hc_{a-1}}^{w}(1-z_{ic_a}^{w}) &\text{if }i = h\\
    -z_{hc_{a-1}}^{w}z_{ic_a}^{w} &\text{otherwise.} \\ 
    \end{cases} \\
    & \forall i\in V, h\in H, w\in W, a\in [2, |w|]
    \end{split}
    \label{eq:nonlinear}
\end{equation}
\normalsize

This contains a multiplication of possible decision variables, that would make the problem non-linear. To solve the problem, we can use some linearization techniques. For this respect, we create the following new variables: $z_{hc_a}^{\prime iw} =z_{hc_{a-1}}^{w} (1-z_{ic_a}^{w} )$, $z_{hc_a}^{\prime\prime iw} =z_{hc_{a-1}}^{w} z_{ic_a}^{w}$. For them to have these values, they have to follow these constraints:

\begin{equation}
    -z_{hc_{a-1}}^{w}+z_{hc_a}^{\prime iw}\le0
    \label{eq:nine}
\end{equation}

\begin{equation}
    -1+z_{ic_a}^{w}+z_{hc_a}^{\prime iw}\le0
    \label{eq:ten}
\end{equation}

\begin{equation}
    z_{hc_{a-1}}^{w}+1-z_{ic_a}^{w}-z_{hc_a}^{\prime iw}\le1
    \label{eq:eleven}
\end{equation}

\begin{equation}
    -z_{hc_{a-1}}^{w}+z_{hc_a}^{\prime\prime iw}\le0
    \label{eq:twelve}
\end{equation}

\begin{equation}
    -z_{ic_a}^{w}+z_{hc_a}^{\prime\prime iw}\le0
    \label{eq:thirteen}
\end{equation}

\begin{equation}
    z_{hc_{a-1}}^{w}+z_{ic_a}^{w}-z_{hc_a}^{\prime\prime iw}\le1
    \label{eq:fourteen}
\end{equation}

(\ref{eq:nonlinear}) can now be rewritten as a linear constraint:

\begin{equation}
    \begin{split}
    \sum_{j\in V} f_{ij}^{hwc_a}-f_{ji}^{hwc_a}=\begin{cases}
    0 & \text{if }i\in S \\
    z_{hc_a}^{\prime iw} &\text{if }i = h\\
    -z_{hc_a}^{\prime\prime iw} &\text{otherwise.} \\ 
    \end{cases} \\
    \forall i\in V, h\in H, w\in W, a\in [2, |w|]
    \end{split}
    \label{eq:fifteen}
\end{equation}

Flow constraints for control flows (i.e. flows of the SDN controllers) are similar. Flow is produced by switches and received by SDN controllers, except for co-located controllers and switches. We consider in-band control traffic, i.e. these control flows are routed through the same network as application traffic, there are not separate links that are specifically for control flows.

\begin{equation}
    \begin{split}
    \sum_{j\in V} cf_{ij}^s - cf_{ji}^s & =\begin{cases}
    0 & \text{if }i\in H \\
    1-y_{si} &\text{if } i = s\\
    -y_{si} &\text{otherwise}
    \end{cases} \\ & \forall i\in V, s\in S
    \label{eq:sixteen}
    \end{split}
\end{equation}

With these flow constraints in place, we also have to account for the maximum link capacity:

\begin{equation}
    \begin{split}
    \sum_{h\in H}\sum_{w\in W}[(\sum_{a = 1}^{|w|}f_{ij}^{hwc_a}I_{c_a})
     +& (f_{ij}^{\prime hw}O_{c_n})] +\sum_{s\in S}[cf_{ij}^s\Omega] <=\theta_{ij} \\ &\forall l_{ij}\in L
    \end{split}
    \label{eq:seventeen}
\end{equation}

An objective function is also required for the model. The role of the objective function is to determine which metric must be optimized by changing the values of the future decision variables. In our case, the objective is the average response time of all workflows. To simplify this, function $SW(i)$ is defined, which is $1$ if $i\in S$ and $0$ otherwise. Formally, the objective function is represented by (\ref{eq:objective}).

\begin{equation}
    \begin{split}
    \sum_{h\in H}\sum_{l_{ij}\in l}\sum_{w\in W}(\sum_{a = 1}^{|w|}(\frac{z_{hc_a}^{w}\xi_{c_a}}{P_h}\\
    + f_{ij}^{hwc_a}\delta_{ij})+f_{ij}^{\prime hw}\delta_{ij}\\
    +SW(j)\sum_{l_{k, m}\in L}cf_{km}^j\delta_{km})
    \end{split}
    \label{eq:objective}
\end{equation}

(\ref{eq:objective}) can be separated into three terms, as shown above. The first term is execution time, and it depends on the workload of each microservice and the power of the host it runs on. The second term is the network latency, which depends on the latency of the links used to transmit information. Finally, the third term is the control latency of the path taken by each workflow.

The final MILP problem is therefore formulated as \emph{minimize (\ref{eq:objective}) subject to (\ref{eq:one}-\ref{eq:eight}) (\ref{eq:nine}-\ref{eq:seventeen})}.
\section{Performance evaluation} \label{sec:evaluation}

In this section, a setup based in the scenario presented in Sec. \ref{subsec:scenario} is shown and tests are performed over this setup to evaluate the performance of \ourprop and compare it with other deployment strategies.

\subsection{Evaluation setup} \label{subsec:setup}

In Sec. \ref{subsec:scenario}, a well-defined scenario that is our basis for evaluating \ourprop, which details are taken from \cite{Sun2018}, is presented. Additionally, we have also estimated values for the parameters that were not reported in the original case study, such as the number of SDN controllers or the length of functionalities. Finally, in \cite{Sun2018}, IIoT devices are considered to be unable to compute. However, in order to evaluate the performance of \ourprop in environments where mist layer devices are available for deployment, we also consider to equip IIoT devices with devices such as Arduino Pro Portenta H7\cite{Arduino2016}, a microcontroller designed for IIoT applications; as well as the unexpensive single-board computer Raspberry Pi Zero\cite{RaspberryPiFoundation2018}, which enables IIoT devices to be a complete computer at a relatively low cost.

Scenarios that allow evaluations of the scalability and performance of \ourprop under different conditions have been tested. Concretely, we consider microservices that took 100, 500 or 1000 MCycles to run; placing between 1 and 4 SDN controllers, with each device requesting between 1 and 4 functionalities, with each functionality workflow being 1, 2, 3 or 6 microservices long and considering the hardware specifications of the IIoT devices stated above, in small (10 IIoT devices, 10 edge servers), medium (25 IIoT devices, 15 edge servers) and large (50 IIoT devices, 25 edge servers) topologies, also taken from \cite{Sun2018}. In general, evaluation is performed by setting a default value for all parameters, varying the values of one or more parameters and testing over the three topologies. The default values used are microservices of 500 MCycles, 1 SDN controller, 2 requests per device, non-computing IIoT devices and 1 microservice long functionalities; values that define our initial case to consider. These values are also shown in Table \ref{tab:eval-params} for quick reference.

\begin{table}[htb]
\caption{Evaluation parameters}
\label{tab:eval-params}
\begin{tabular}{|l|l|l|}
\hline
\textbf{Parameter} & \textbf{Values} & \textbf{Unit} \\ \hline
Microservice workload & 100, 500, 1000 & MCycles \\ \hline
SDN controllers & 1, 2, 3, 4 & Controllers \\ \hline
Requests per device & 1, 2, 3, 4 & Requests \\ \hline
Functionality length & 1, 2, 3, 6 & Microservices \\ \hline
Topology size & 20, 40, 80 & Nodes \\ \hline
IIoT device hardware &
\begin{tabular}{c}
Non-computing, \\Arduino Pro Portenta H7, \\ Raspberry Pi Zero
\end{tabular}
 &  \\ \hline
\end{tabular}
\end{table}

The objectives of our evaluation are multiple. The validity of \ourprop is to be evaluated, while performing tests to show the scalability of \ourprop as IIoT devices request more functionalities, as well as its computational scalability. This is crucial, since the company may not deem worthy to invest in the transformation if response time increases heavily when the system grows. Another key aspect of the evaluation is the trade-off between latency and response time, in order to observe if \ourprop picks a solution with higher latency if the reduction in execution time of said solution brings an overall lower response time. Finally, \ourprop is compared against alternative deployment strategies to evaluate the response time decrease obtained by \ourprop.

\subsection{Evaluation results} \label{subsec:results}

The results shown in this section are acquired by combining the solutions obtained by \ourprop and the values for the parameters previously discussed, calculating the values of the different metrics that are shown to analyze the expected behaviour of \ourprop under different conditions.

\begin{figure*}
    \begin{subfigure}[b]{0.32\textwidth}
        \centering
        \includegraphics[width=1.05\textwidth]{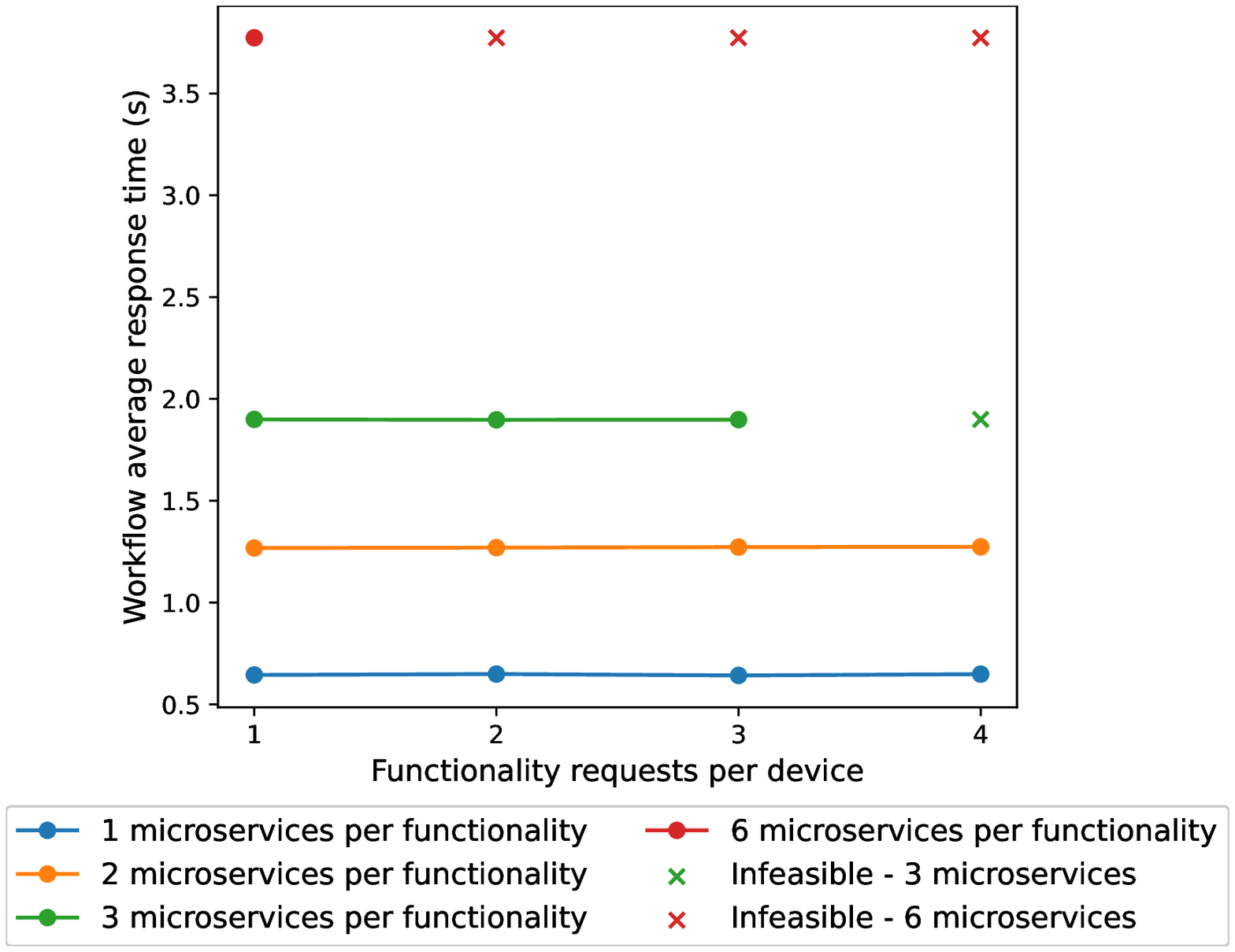}
        \caption{Small topology.}
        \label{subfig:small}
    \end{subfigure}
    \hfill
    \begin{subfigure}[b]{0.32\textwidth}
        \centering
        \includegraphics[width=0.98\textwidth]{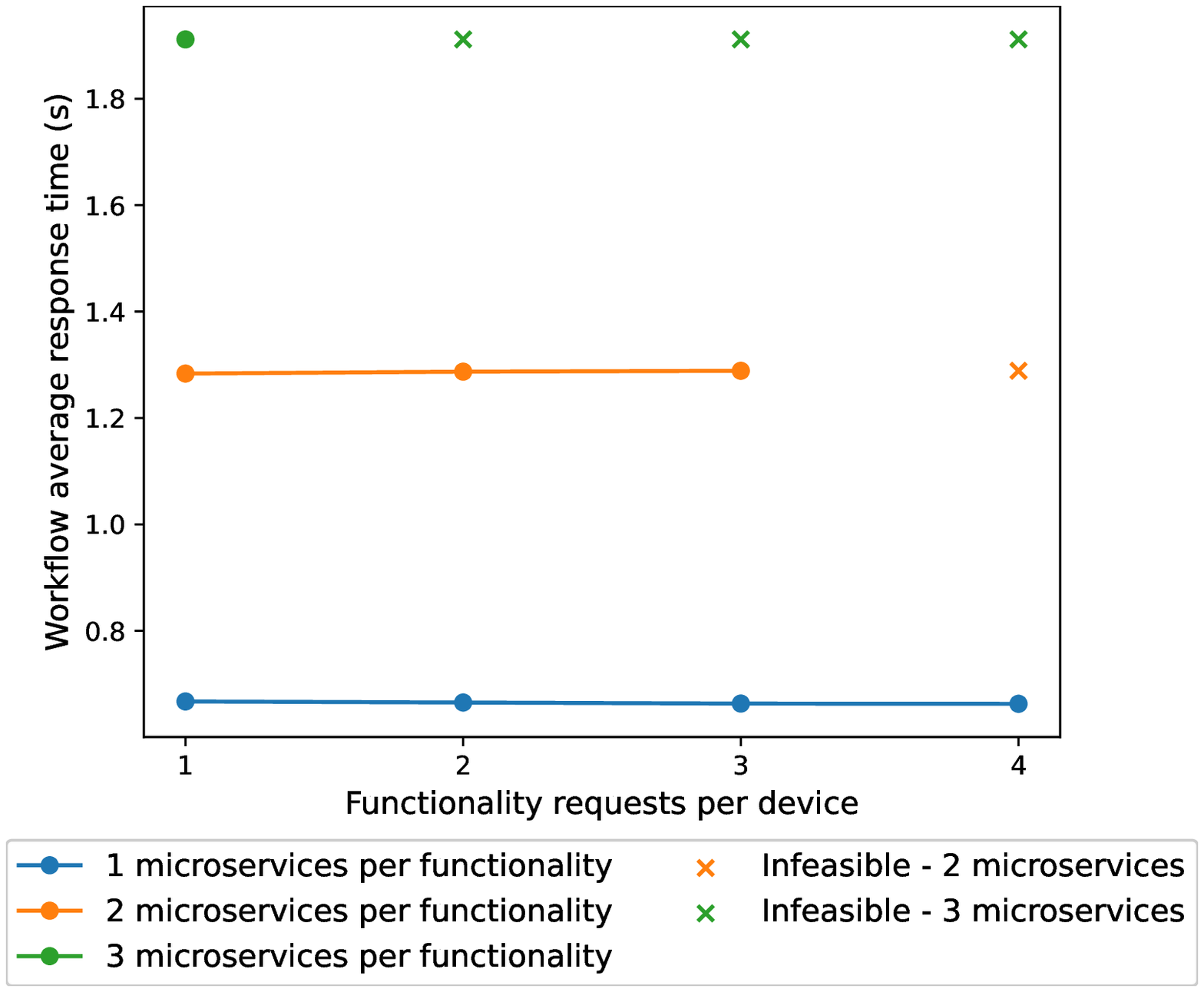}
        \caption{Medium topology.}
        \label{subfig:medium}
    \end{subfigure}
    \hfill
    \begin{subfigure}[b]{0.32\textwidth}
        \centering
        \includegraphics[width=0.98\textwidth]{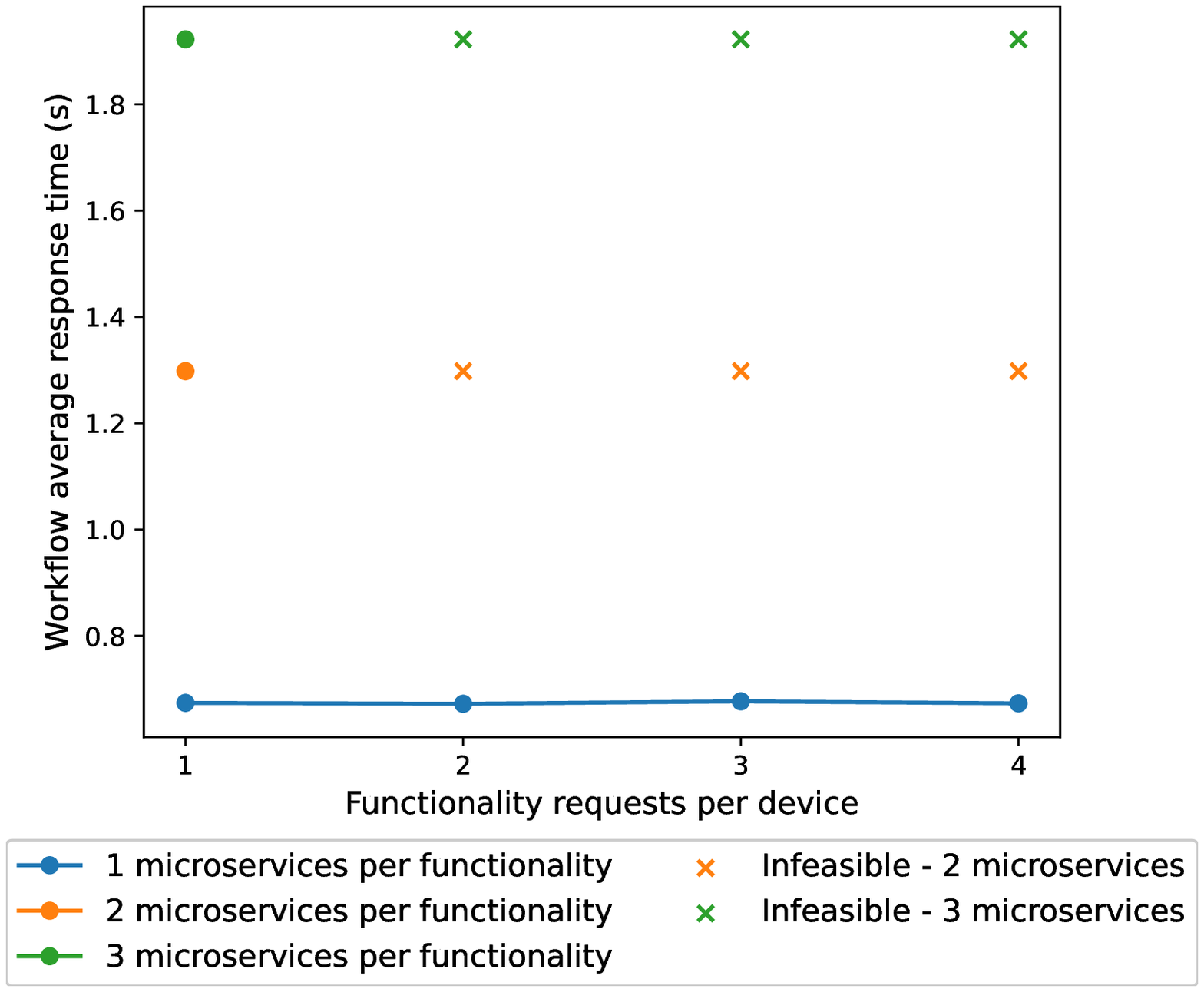}
        \caption{Large topology.}
        \label{subfig:large}
    \end{subfigure}
    \caption{Request scalability analysis.}
    \label{fig:scalab}
\end{figure*}

In Fig. \ref{fig:scalab}, the scalability of the solution is evaluated by testing the deployment of applications in all three topologies with an increasing number of microservices, as well as by increasing the amount of functionalities requested. These tests are performed with a single SDN controller, microservices of 500 MCycles and non-computing IIoT devices. There are four major conclusions to draw from these results: first, we can see that the solution is scalable. When requests per device increase, response time stays stable. This implies that the solution has good scaling, and therefore is able to maintain optimal response time. However, the second conclusion is that the solution can only scale as long as the architecture has enough resources to deploy the expected applications. There are points labeled as \emph{Infeasible}. These points are the estimated positions for parameter values that required more resources (e.g. memory or networking capacity) than the architecture had available, and could not be solved because of it. The third conclusion is that functionality length (in microservices) is very relevant to response time, mainly because a functionality workflow with more microservices implies a heavier computation load (e.g. a two microservice long functionality has twice the computation load of a single microservice long functionality), but also because there is an extra cost in latency if these microservices are not executed in the same host. Finally, as the IIoT devices per edge servers ratio rises in larger topologies, less microservices per request can be supported, but the results are very similar in all topologies. This also applies to the rest of the performed analyses, and therefore, from now on, results from a single topology are shown, while the rest of the results are provided as additional content. Therefore, \ourprop provides scalable solutions to deploy an application, as long as the capabilities of the architecture are enough to withstand said application.

\begin{figure}
    \centering
    \includegraphics[width=0.75\columnwidth]{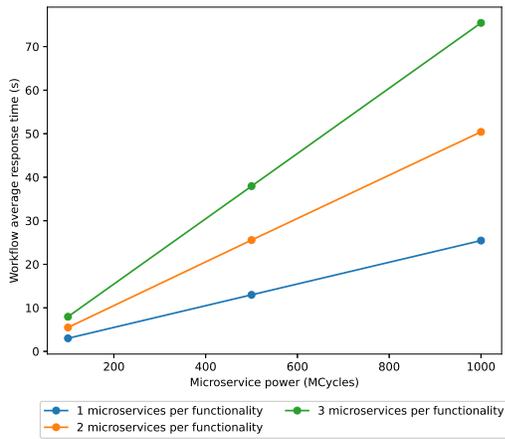}
    \caption{Computational scalability analysis in the small topology.}
    \label{fig:cycles}
\end{figure}

In Fig. \ref{fig:cycles}, the scalability of the application when the computational load changes is analyzed in the small topology. These tests are performed with a single SDN controller, 2 requests per device and non-computing IIoT devices. Firstly, the slope is more steep when functionalities are longer, which implies that, the more microservices are executed on a functionality workflow, the longer it takes to execute said workflow, responding to the nature of this service composition (e.g. a single microservice long functionality with 1 GCycle microservices has to execute 1GCycle, whereas a two microservice long functionality would have to execute 2 GCycles). However, it is not an exact measurement: while a single microservice long functionality with 1 GCycle microservices and a two microservices long functionality with 500 MCycles microservices have the same computational workload, the response time of the shorter functionality is slightly lower: concretely, it is 11 ms lower. This is because of the communication latency, since there is a communication delay between each of the microservices in the functionality workflow that does not exist when there is a single microservice. However, while shorter functionalities have a smaller response time, longer functionalities can be parallelized and its microservices can be used by other requests, thus compensating that slight overhead. Overall, \ourprop provides a solution that is able to minimize the effect of latency in response time.

\begin{figure}
    \centering
    \includegraphics[width=0.8\columnwidth]{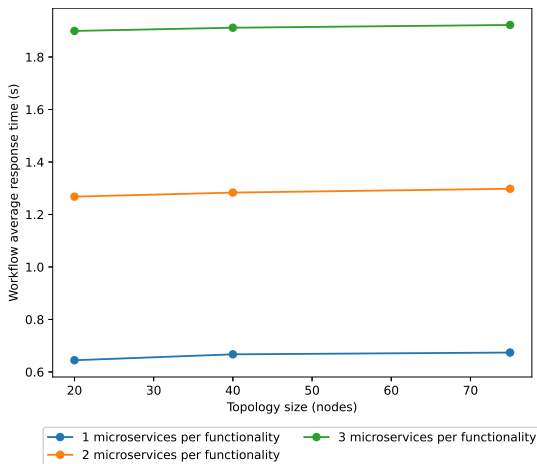}
    \caption{Comparison of response time in different topologies.}
    \label{fig:topocomp}
\end{figure}

In Fig. \ref{fig:topocomp}, the scalability of the application in different topologies is tested. These tests are performed with a single SDN controller, 2 requests per device, microservices of 500 MCycles and non-computing IIoT devices. These results further prove the scalability of the solutions provided by \ourprop, with only a slight increase in response time as the topology size doubles or even triplicates the original size. The main conclusion to draw from them is that the IIoT application that is to be implemented as explained in Sec. \ref{sec:motivation} can be scaled into a larger network and the increase in response time experienced after this scaling is not very large, and thus there will not be an important QoS decrease when the system grows.

\begin{figure}
    \centering
    \includegraphics[width=0.8\columnwidth]{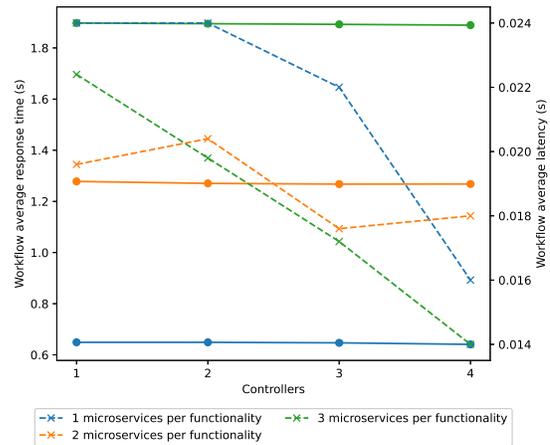}
    \caption{Latency and controller analysis in the small topology. Solid lines show latency, dashed lines show response time.}
    \label{fig:latcont}
\end{figure}

In Fig. \ref{fig:latcont}, the effect of adding SDN controllers on latency and response time is analyzed in the small topology. Solid lines show response time and refer to the leftmost Y-axis, while dashed lines show latency and refer to the rightmost Y-axis. These tests are performed with 2 requests per device, microservices of 500 MCycles and non-computing IIoT devices. In short functionalities (1 microservice per functionality), latency does not decrease steeply when a single controller is added, but it gets steeper as more controllers are added. If the functionality is longer (2 microservices per functionality), latency even rises if there is an even number of controllers, but there's a steep decrease when odd numbers of controllers are considered. On even longer functionalities (3 microservices per functionality), latency decreases almost linearly when controllers increase. However, response time decreases slightly and steadily as controllers are added in every case. The essential conclusion is that the effects of adding SDN controllers over latency heavily depend on how traffic flows are steered, further proving that the computing and networking dimension affect the QoS of each other. Furthermore, latency rises in the case with functionalities of length 2 as a result of a trade-off, in which \ourprop esteemed worthy to choose a deployment with higher latency because it would decrease execution time in such a way that overall response time is minimized, something that would not be as simple if both dimensions were considered separately. Thus, \ourprop is able to find the solution to this trade-off between execution time and latency, which not only enables for a smart computation distribution scheme, offloading computation when latency is low enough for it to be worth it, but also enables for a smart controller placement that considers and complements these offloading decisions.

\begin{figure}
    \centering
    \includegraphics[width=0.75\columnwidth]{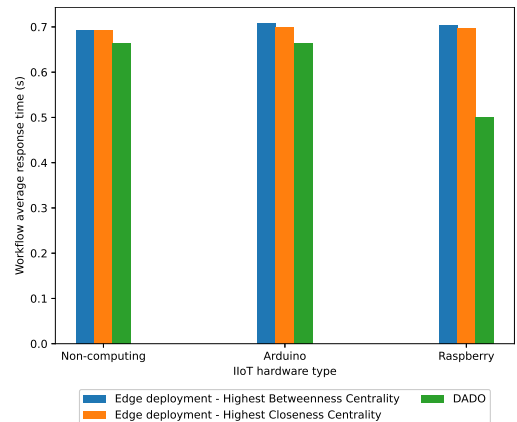}
    \caption{Performance comparison of \ourprop in the medium topology.}
    \label{fig:perfcomparison}
\end{figure}

In Fig. \ref{fig:perfcomparison}, the performance of \ourprop is compared against other solutions in the medium topology. Concretely, \ourprop is compared with deploying the application directly in the edge and placing the SDN controllers in the nodes with the Highest Betweenness Centrality (HBC) or Highest Closeness Centrality (HCC). Edge deployments are performed by matching microservices with edge servers in a round-robin fashion, making sure that the total memory of edge servers is never surpassed. In both of these cases, the routing is still optimized through the usage of the formulation proposed by \ourprop. These tests are performed with a single SDN controller, 2 requests per device, 1 microservice long functionalities and microservices of 500 MCycles. The key conclusion to this analysis is that \ourprop provides better results than the alternative solutions, speeding up execution time by up to a 28.97\% on average. \ourprop achieves a better deployment mainly because of its capabilities to deploy in a hybrid manner: while other approaches only solve the DCDP on the fog or edge layer, \ourprop will use all the available layers and combinations of them to optimize response time, automatically determining the best layer to offer each microservice. This is clearly shown on the Arduino hardware: \ourprop refuses to deploy in Arduino devices because, although it is possible to deploy some microservices there, execution would be so slow that response time would worsen.

\begin{table}
\caption{Multivariate analysis of response time.}
\label{tab:multivar}
\begin{tabular}{l|l|l|l|}
\cline{2-4}
 & \textbf{Coefficient} & \textbf{Std. Error} & \textbf{P-value} \\ \hline
\multicolumn{1}{|l|}{\textbf{Intercept}} & 1.2276 & 0.272 & $\mathbf{2.3\cdot10^{-5}}$ \\ \hline
\multicolumn{1}{|l|}{\textbf{Topology size}} & 0.0002 & 0.003 & 0.935 \\ \hline
\multicolumn{1}{|l|}{\textbf{SDN controllers}} & -0.0163 & 0.052 & 0.757 \\ \hline
\multicolumn{1}{|l|}{\textbf{Requests per device}} & -0.0051 & 0.069 & 0.941 \\ \hline
\multicolumn{1}{|l|}{\textbf{Functionality length}} & 0.6503 & 0.046 & $\mathbf{2.5\cdot10^{-23}}$ \\ \hline
\multicolumn{1}{|l|}{\textbf{Cycles per microservice}} & -1.1725 & 0.091 & $\mathbf{3.3\cdot10^{-21}}$ \\ \hline
\multicolumn{1}{|l|}{\textbf{IIoT device hardware}} & -0.3071 & 0.171 & 0.077 \\ \hline
\end{tabular}
\end{table}

In order to statistically validate the results descrived above, a multivariate analysis has been performed to evaluate the impact of a set of parameters over response time, which is set as the dependent variables. Table \ref{tab:multivar} shows the results of this analysis. The independent variables are the topology size (number of nodes), the number of SDN controllers, the number of requests per device, the length of functionalities, the number of cycles per microservice and the type of IIoT hardware as a binary variable (0 for non-computing and Arduino, 1 for Raspberry Pi). The $R^2$ coefficient of this analysis is of $0.841$. The main conclusion of this analysis is that the size of the application, mainly expressed through the length of the functionalities and the cycles per microservice, is the most statistically significant variable for response time. However, they are bound to be in balance: to reduce the cycles per microservice, more microservices per functionality are required, which raises their length and likely adds some overhead. To shorten the functionality, different microservices can be merged into a single one, but that makes these microservices heavier (i.e. more cycles per microservice) and they often require more memory as well. Other interesting conclusions can be extracted from the coefficients of each variable, which explain how response time rises or lowers as they are modified (e.g. adding a controller makes response time be reduced in 0.0163 s).

Since the Edge-SDN Deployment Problem is a combination of the DCDP and the CPP, both of which have been proven to be NP-hard\cite{Das2019,Brogi2020}, and by definition the combination of any given NP-hard problem with another problem must also be NP-hard, the Edge-SDN Deployment Problem is also NP-hard. This NP-hardness is reflected on the limitations that DADO has. The MILP solution of DADO can only be feasibly applied to infrastructures with under 300 nodes. In larger infrastructures, the formulation states that it needs to allocate nearly 18 Exabytes of memory.
\section{Related work} \label{sec:related}

In an attempt to achieve a short response time in different applications, there is an ongoing research on edge computing paradigms, both to apply these paradigms and to standarize them. Yousefpour \emph{et al.} \cite{Yousefpour2019} have surveyed the different cloud and edge computing paradigms and provide insights on the similarities and differences between them. While this paper evaluated \ourprop on a hybrid edge and mist computing scenario, \ourprop is designed to optimize deployments that make use of other edge computing paradigms as well, such as fog computing. Bellavista \emph{et al.} \cite{Bellavista2019} have surveyed different proposals that leverage the fog computing paradigm for its use on IoT applications as an approach to support the strict response time requirements of some of these applications. However, these proposals are mostly different platforms that make use of fog computing and provide services such as communication, security or resource virtualization that simplify the deployment of IoT applications in fog environments; but they mostly do not provide a solution to optimize deployment when using said platforms.

The problem of optimizing the deployment of microservices to an edge computing infrastructure, focusing on the computing dimension, is the DCDP. The term DCDP was coined by Choudhury \emph{et al.} \cite{Choudhury2019}, who have solved the DCDP in mist computing environments. The main idea behind this DCDP instance is to replicate cloud services of applications in smartphone and IoT devices to enhance their response time, to then allow nearby devices to consume the replicated services within an ad-hoc network. This instance of the DCDP focuses on optimizing the resources used for replication in the mist layer while delivering an appropriate response time, instead of focusing on other edge computing environments, delivering an optimal response time or optimizing the networking dimension as well like \ourprop does. Carrega \emph{et al.} \cite{Carrega2017} propose another instance for the DCDP that allows an application with an MSA to be optimally deployed in an edge infrastructure taking into account the dependencies between microservices. The architecture of their framework is thoroughly explained, but there is no public formulation to the DCDP they solve, and they do not take into account the possibility of optimizing the network layer. Sun \emph{et al.} \cite{Sun2018} propose a double-auction heuristic scheme for optimizing the deployment of IIoT applications in an edge environment. In it, the response time offered by edge servers and required by services are assessed as prices, and an auction-based algorithm is leveraged to optimize the deployment. However, this work does not optimize response time and, instead, tries to optimize the amount of services that can be successfully deployed. A large amount of proposals for solutions to the DCDP are surveyed in \cite{Brogi2020} using a variety of techniques such as MILP, heuristics or game theory. Although some of them partially optimize the network latency through routing, as \ourprop does as well, none of them consider to optimize the latency of the network through SDN controller placement.

Edge computing paradigms, while supporting QoS requirements that are complicated to support with cloud computing, also have challenges that need to be accounted for, as the previously discussed service discovery problem. \cite{Baktir2017} presents a proposal for applying SDN as a solution to these challenges that is transparent to the different hosts involved in the network. Other research efforts, such as \cite{Lv2019} or \cite{Huang2019}, solve the DCDP with an underlying SDN network. Although these solutions are also built to support IoT applications, they do not consider the effects on latency of SDN controller placement.

On the networking plane, the SDN CPP is a well-known problem that has an important impact on the latency provided by the network. Zhang \emph{et al.} \cite{Zhang2018} solved a multi-objective version of the CPP through an Adaptive Bacterial Foraging Optimization algorithm, taking into account controller load balancing, reliability and latency; as well as allowing the user of the solution give different weighs to each of the objectives. However, this solution does not consider that control loads come from routing the data from the computing plane. \cite{ulHuque2015} adds the idea of dynamic flows to the CPP, and therefore does not only solve the CPP for pre-defined, static flows, but allows this solution to be varied when the flows vary (e.g. because of changes in routing or traffic). Although this is a relevant contribution to the CPP, the relationship between the networking and computing plane is still unconsidered. Another kind of approach to the CPP is the dynamic provisioning CPP \cite{Bari2013}. Instead of optimally placing SDN controllers, the dynamic provisioning CPP assumes every SDN switch also has an SDN controller and instead focuses on managing which controllers are on and which ones should be off. This is a completely different approach to the CPP that is not yet considered by \ourprop. Singh \emph{et al.} \cite{Singh2020} not only solved the CPP with a Varna-based heuristic approach, but also classified the CPP into 12 kinds, depending on the SDN network features that are considered. \ourprop is a type 4, uncapacitated CPP, but also adds routing capabilities that are unconsidered in this classification; as well as it relates the computing and networking planes. Finally, \cite{Das2019} surveys a large amount of proposals for CPP solutions using similar techniques to the solutions to the DCDP, but also failing to integrate the computing plane into them like \ourprop does.

While both the DCDP and CPP problems are well-known, to the best of our knowledge, no other works integrate the DCDP and the CPP into a single, cohesive problem like \ourprop does; nor do they take into account the relationship and influences between the two problems. Therefore, the contribution of \ourprop is an integration of the CPP and the DCDP to provide an optimal response time for IIoT applications.
\section{Conclusions and future work} \label{sec:conclusions}

As the interest of applying IoT to intensive domains such as Industry or Healthcare \cite{Xu2018, Greco2020} rises, achieving optimal response time in intensive IoT applications becomes crucial, and so becomes leveraging edge computing and SDN in these infrastructure, as well as optimizing them to get the required response time. This work defines and formalizes the problem of optimizing an edge computing and SDN infrastructure for QoS-strict IoT applications with an MSA, merging the optimization efforts from both the computing and the networking dimension. By optimally distributing microservices between the nodes, optimally placing SDN controllers and taking into account the mutual influences between both dimensions, optimal decisions are made and sub-optimal solutions are avoided. This work also proposes a framework, \ourprop, that is able to solve this joint problem, and evaluates \ourprop over an IIoT scenario under different conditions, showing the scalability and trade-off optimizing capabilities of the solution, as well as its performance compared to alternative solutions. \ourprop reduces response time by up to 28.97\% by optimizing the deployment and allowing for hybrid (e.g. edge and mist layer) edge computing deployments.

In the future, we consider to address some of \ourprop's limitations. First, the MILP version of \ourprop is, computationally, very complex, which reflects on its time and memory consumption. In the future, we consider to develop heuristics that allow \ourprop to be applied to infrastructures larger than 300 nodes while still finding near-optimal solutions. Furthermore, these heuristics can be used to execute \ourprop periodically in short loops, enabling \ourprop to maintain the QoS when the environment changes by adapting the deployment to new conditions over time. Another limitation is the case study, that only considers edge and mist deployments. \ourprop is to be tested on larger scenarios with multiple layers, including fog or cloud nodes, in future works. Finally, \ourprop can only enhance response time. In the future, we consider to expand \ourprop to consider other QoS features, such as reliability, as well as to combine these QoS features into a multi-objective version of \ourprop that is able to optimize features such as reliability along with response time.

\bibliographystyle{IEEEtran}
\bibliography{biblio}

\begin{IEEEbiography}[{\includegraphics[width=1in,height=1.25in,clip,keepaspectratio]{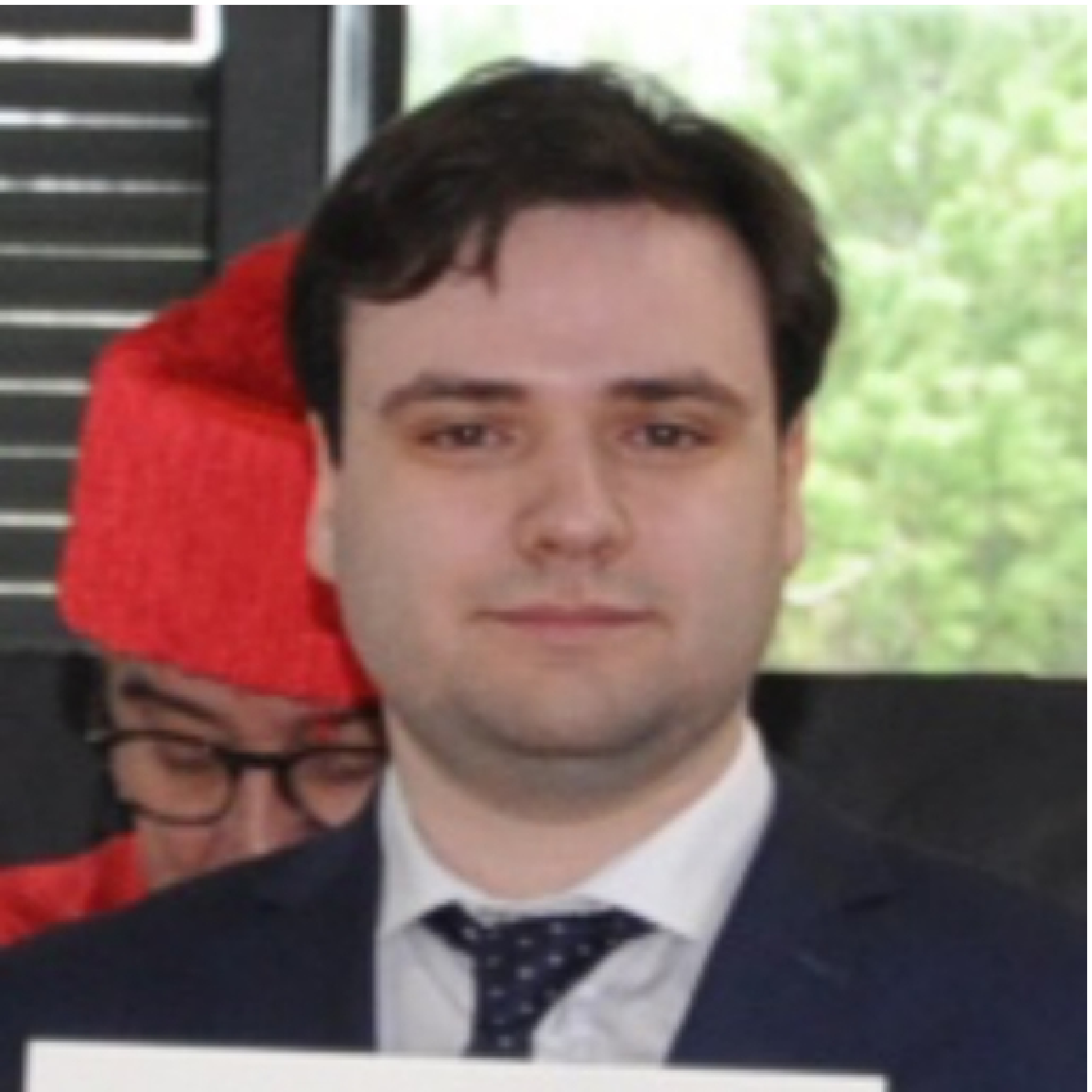}}]{Juan Luis Herrera} received the Bachelor degree in software engineering from the University of Extremadura in 2019. He is a researcher at University of Extremadura's Computer Science and Communications Engineering Department. His main research interests include IoT, fog computing and SDN.
\end{IEEEbiography}

\begin{IEEEbiography}[{\includegraphics[width=1in,height=1.25in,clip,keepaspectratio]{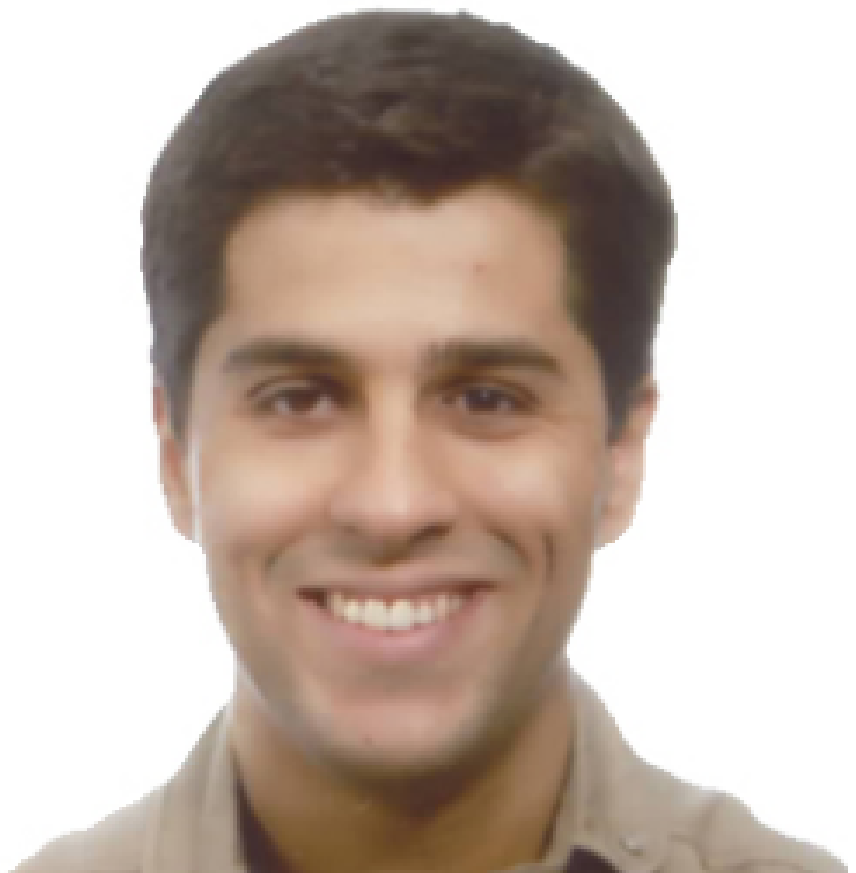}}]{Jaime Galán-Jiménez} received the Ph.D. degree in computer science and communications from the University of Extremadura in 2014. He is currently with the Computer Science and Communications Engineering Department, University of Extremadura, as an Assistant Professor. His main research interests are Software-Defined Networks, 5G networks planning and design, and mobile ad-hoc networks.
\end{IEEEbiography}

\begin{IEEEbiography}[{\includegraphics[width=1in,height=1.25in,clip,keepaspectratio]{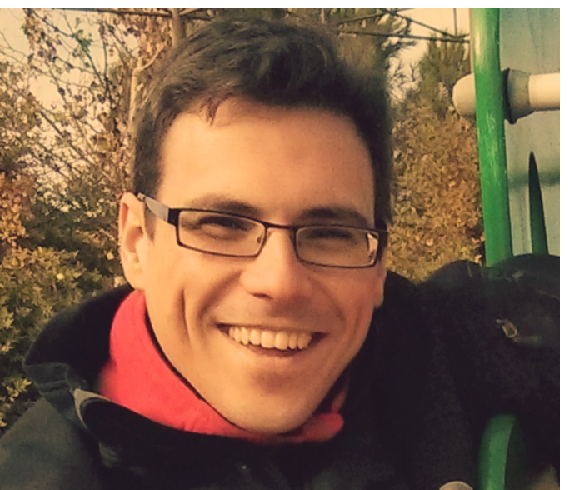}}]{Javier Berrocal} (IEEE Memeber) is a cofounder of Gloin. His main research interests are software architectures, mobile computing, edge and fog computing. Berrocal has a PhD in computer science from the University of Extremadura, where he's currently an associate professor.
\end{IEEEbiography}

\begin{IEEEbiography}[{\includegraphics[width=1in,height=1.25in,clip,keepaspectratio]{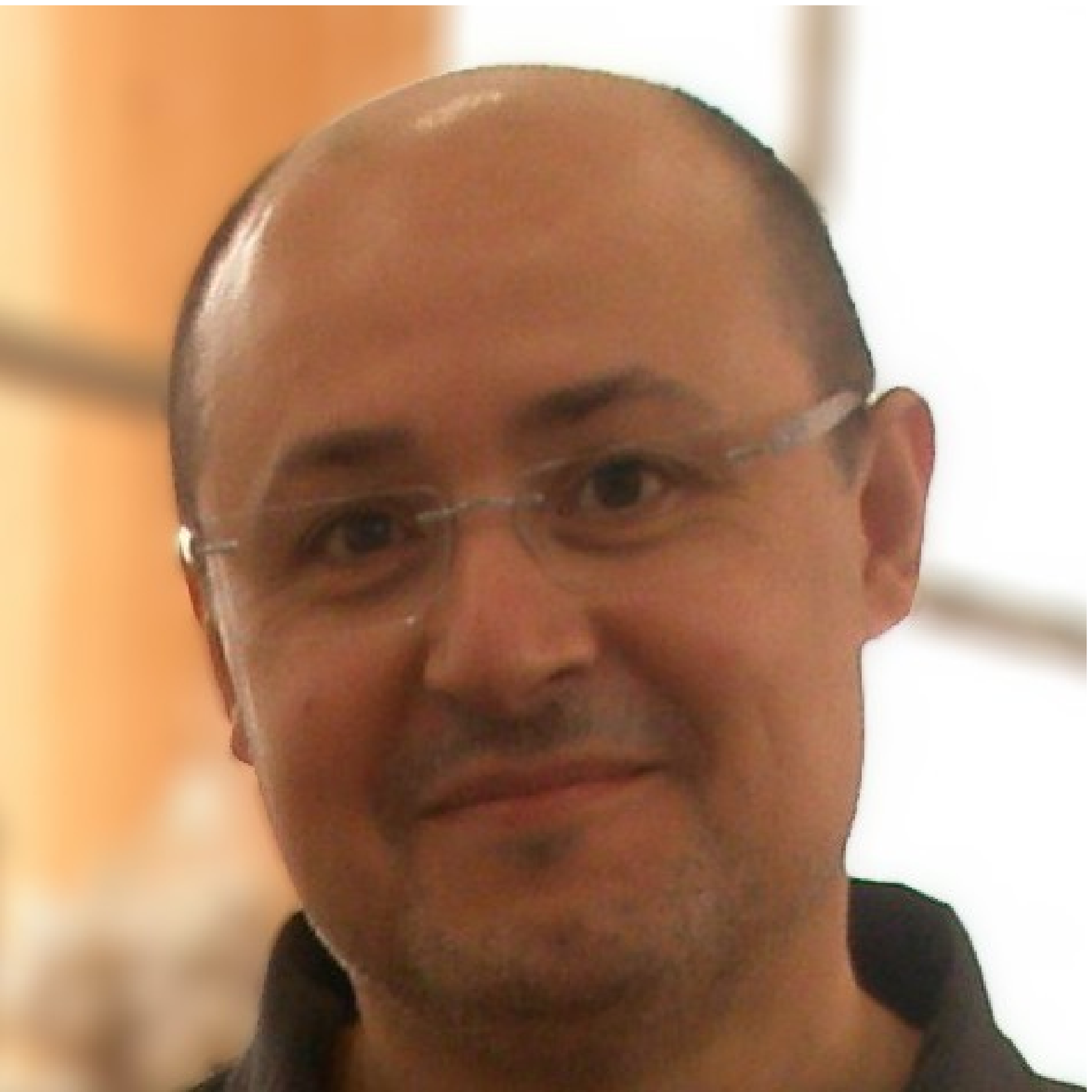}}]{Juan M. Murillo} (IEEE Member) is a cofounder of Gloin and a full professor at the University of Extremadura. Murillo is full professor at the the University of Extremadura. His research interests include software architectures, mobile computing, and cloud computing.
\end{IEEEbiography}

\end{document}